\documentclass[12pt]{article}

\renewcommand{\d}{{\rm d}}
\newcommand{\dotOm}{\dot{\Omega}}

\newcommand{\dagOm}{\Omega^\dagger}
\newcommand{\dagu}{u^\dagger}
\newcommand{\dotu}{\dot{u}}
\newcommand{\vf}{\mbox{\boldmath$f$}}
\newcommand{\vn}{\mbox{\boldmath$n$}}
\newcommand{\vb}{\mbox{\boldmath$b$}}
\newcommand{\vl}{\mbox{\boldmath$l$}}
\newcommand{\ve}{\mbox{\boldmath$e$}}
\newcommand{\dotvn}{\dot{\vn}}
\newcommand{\dotve}{\dot{\ve}}

\newcommand{\prvn}{\vn^\prime}
\newcommand{\prve}{\ve^\prime}
\newcommand{\prdotvn}{\dotvn^\prime}

\newcommand{\pprvn}{\vn^{\prime\prime}}
\newcommand{\pprve}{\ve^{\prime\prime}}
\newcommand{\pprdotvn}{\dotvn^{\prime\prime}}
\newcommand{\pprdotve}{\dotve^{\prime\prime}}

\newcommand{\llangle}{{\overbrace{\vl_1 \vl}}_{\! 2}}
\newcommand{\3}{{31^+}}
\newcommand{\0}{{02}}
\newcommand{\nllangle}{{\overbrace{\vl_\3 \vl}}_{\0}}

\newcommand{\tvn}{\tilde{\vn}}
\newcommand{\dottvn}{\dot{\tvn}}
\newcommand{\vvarep}{\mbox{\boldmath$\varepsilon$}}

\renewcommand{\cosh}{{\rm ch}}
\renewcommand{\sinh}{{\rm sh}}

\begin{document}

\title{On area spectrum in the Faddeev gravity}
\author{V.M. Khatsymovsky \\
 {\em Budker Institute of Nuclear Physics} \\ {\em of Siberian Branch Russian Academy of Sciences} \\ {\em
 Novosibirsk,
 630090,
 Russia}
\\ {\em E-mail address: khatsym@inp.nsk.su}}
\date{}
\maketitle
\begin{abstract}
We consider Faddeev formulation of gravity, in which the metric is bilinear of $d = 10$ 4-vector fields. A unique feature of this formulation is that the action remains finite for the discontinuous fields (although continuity is recovered on the equations of motion). This means that the spacetime can be decomposed into the 4-simplices virtually not coinciding on their common faces, that is, independent. This allows, in particular, to consider a surface as consisting of a set of virtually independent elementary pieces (2-simplices). Then the spectrum of surface area is the sum of the spectra of independent elementary areas. We use connection representation of the Faddeev action for the piecewise flat (simplicial) manifold earlier proposed in our work. The spectrum of elementary areas is the spectrum of the field bilinears which are canonically conjugate to the orthogonal connection matrices. We find that the elementary area spectrum is proportional to the Barbero-Immirzi parameter $\gamma$ in the Faddeev gravity and is similar to the spectrum of the angular momentum in the space with the dimension $d - 2$. Knowing this spectrum allows to estimate statistical black hole entropy. Requiring that this entropy coincide with the Bekenstein-Hawking entropy gives the equation, known in the literature. This equation allows to estimate $\gamma$ for arbitrary $d$, in particular, $\gamma = 0.39...$ for genuine $d = 10$.
\end{abstract}

keywords: Faddeev gravity; piecewise flat spacetime; connection; area spectrum


MSC classes: 83C27; 53C05

\section{ Introduction}

Recently Faddeev has proposed \cite{Fad} a new formulation of Einstein's gravity. In this formulation the metric is composed of vector fields or the tetrad of the ten-dimensional fields, $f^A_\lambda$, where $\lambda$ run over the four values, usually taken as 1, 2, 3, 4 in the case of the Euclidean metric signature or 0, 1, 2, 3 for the considered in the present paper Minkowsky signature. Correspondingly, the index $A$ is Euclidean/Minkowsky index of the external flat space of dimension $d = 10$ (next we discuss the case of an arbitrary $d$).

The metric tensor in the Faddeev formulation is thus
\begin{equation}                                                            
g_{\lambda \mu} = f^A_\lambda f_{\mu A}.
\end{equation}

\noindent The action is functional of the field $f^A_\lambda$,
\begin{equation}\label{S}                                                   
S = \frac{1}{16 \pi G} \int \Pi^{AB} (f^\lambda_{A, \lambda} f^\mu_{B, \mu} - f^\lambda_{A, \mu} f^\mu_{B, \lambda}) \sqrt {g} \d^4 x.
\end{equation}

\noindent Here $\Pi_{AB} = \delta_{AB} - f^\lambda_A f_{\lambda B}$, $f^\lambda_{A, \mu} \equiv \partial_\mu f^\lambda_A$. Upon partial use of the equations of motion for $f^A_\lambda$ the Eq. (\ref{S}) becomes the Einstein action.

A feature of the Faddeev action (\ref{S}) is that it remains finite for the fields $f^A_\lambda$ (and therefore the metrics $g_{\lambda \mu}$) discontinuous along a coordinate $x^\mu$, since the action does not contain any of the squares of derivatives. In this regard, the theory is different from any of the usual field theories, including the usual Einstein theory. Despite the fact that this discontinuity does not survive on the equations of motion in classical framework (where, remind, this theory reduces to Einstein's one), the discontinuous metric is possible in the Faddeev gravity virtually in quantum framework.

In the minisuperspace formulation of the Faddeev gravity on the piecewise flat spacetime \cite{II} the field $f^A_\lambda$ is piecewise constant. This field is constant in the 4-simplices that make up the piecewise-flat manifold and, in turn, determine the length of their edges. Above possibility of the virtual discontinuity of $f^A_\lambda$ means that edge lengths of the neighboring 4-simplices are considered as independent variables. In particular, these 4-simplices do not fit on their common faces. This independence of the different 4-simplices means considerable methodological simplification for it saves us from having to impose additional geometrical constraints requiring uniqueness of the length of the same edge defined in the different 4-simplices.

The immediate task that is simplified due to the independence of neighboring 4-simplices is to analyze the spectrum of surface area. To this end, we consider the surface as consisting of a set of virtually independent elementary pieces (2-simplices or triangles). Spectrum of the total surface area is the sum of the spectra of independent elementary areas.

The elementary area tensor is the bilinear in the fundamental discrete field variable $f^A_\lambda$. The most direct way to perform the canonical quantization on the basis of the Hamiltonian formalism is achieved by using the connection representation of the Faddeev gravity \cite{I} or, in the present context, its discrete analog \cite{II}. Roughly speaking, elementary area is canonically conjugate to the connection variable which is an orthogonal matrix in the $d$-dimensional external spacetime. This fact leads to quantization of the elementary area in qualitative analogy with the way it happens with the quantization of angular momentum that is canonically conjugate to an orthogonal matrix of rotation in three-dimensional space. Namely, this quantization follows from the requirement that the wave function be single-valued w. r. t. the angle variables on which it depends. Specifically, the elementary area turns out to be quantized as the momentum in the space with the dimension $d - 2$.

The quantization of the surface area was discussed in the continuum theory as well, namely, using Ashtekar variables as early as in the work \cite{Ash}. But there is another principle. They choose a specific expression for the operator of the surface area. The nontriviality of the procedure for determining such an operator is the need to define the product of field operators (namely, the tetrad) at one point. Way to make sense of this expression is the point splitting regularization. To preserve gauge invariance in such a splitting the path ordered exponents of the connection field operator are introduced. At the same time, the tetrad and the connection are canonically conjugate and have non-trivial commutators with each other. This leads to the fact that evaluation of the area operator on certain set of states in the Hilbert space of states (loop states) gives a discrete set of values. In overall, the mentioned regularization issue together with the local gauge symmetry requirement results in the nontrivial discrete set of values of the area operator.

The surface area operators in terms of the {\it discrete} Ashtekar type variables were considered in Ref. \cite{Loll}.

The spectrum of the surface area plays an important role in the black hole physics, in particular, in reproducing proportionality of the black hole entropy to its horizon area (Bekenstein-Hawking relation 
) by statistical method.
If known, the spectrum of horizon area can be used to calculate the black hole entropy and find the condition that this entropy coincides with the Bekenstein-Hawking entropy. In Ref \cite{ABCK} (see also Refs. \cite{LewDom,Mei}) this condition has allowed to find the Barbero-Immirzi parameter $\gamma$ \cite{Barb,Imm} to which the spectrum turns out to be proportional, for a simplified choice of the form of spectrum of the horizon area compared to the spectrum of the generic surface area. The Barbero-Immirzi parameter determines a term which can be added to the Cartan-Weyl form of the Hilbert-Einstein action and which vanishes on the equations of motion for connections \cite{Holst,Fat} thus leading to the same Einstein action in terms of purely metric. Also there is a requirement of the so-called holographic bound principle for the entropy of any spherical nonrotating system including black hole \cite{Bek2,Tho,Sus}. To meet this requirement, in Refs \cite{Khr1,Glo,Khr2,Cor} the formula for the spectrum of the horizon area was chosen to coincide with the general formula for the spectrum of the surface area, and corresponding value of $\gamma$ found.

The concepts of the Lagrangian and the canonical formalism suggest that one of the coordinates (time) is continuous. The rest of the article begins with the transition to the limit of continuous time in a fully discrete action \cite{II} in order to find the kinetic part of the Lagrangian of the Faddeev discrete gravity (symbolically, $p\dot{q}$). Through analysis of this kinetic part, we find the spectrum of the elementary area, arising under the canonical quantization. The spectrum obtained is used to estimate the parameter $\gamma$ using the equation, known in the literature, which expresses the condition that the statistical entropy of a black hole is equal to the Bekenstein-Hawking its entropy. Thus, we can find $\gamma$ for the Faddeev gravity for various dimensions $d$ for the external spacetime, in particular, $\gamma = 0.39...$ for the genuine $d = 10$.

\section{ Kinetic term of the Lagrangian}

We consider the Faddeev action for the manifold composed of hypercubes \cite{II},
\begin{eqnarray}\label{S-discr}                                             
S^{\rm discr} & = & \frac{1}{8 \pi G} \sum_{\rm sites} \sum_{\lambda, \mu} \frac{\sqrt{ (f^\lambda )^2 (f^\mu )^2 - (f^\lambda f^\mu )^2}}{2 \sqrt{-\det \| f^\lambda f^\mu \|}} \arcsin \left[ \frac{f^\lambda_A f^\mu_B - f^\mu_A f^\lambda_B}{2 \sqrt{ (f^\lambda )^2 (f^\mu )^2 - (f^\lambda f^\mu )^2}} \right. \nonumber \\ & & \left. \cdot R^{AB}_{\lambda\mu} (\Omega ) \right] + \sum_{\rm sites} \sum_{\lambda, \mu, \nu} \Lambda^\lambda_{[\mu \nu]} \Omega^{AB}_\lambda (f^\mu_A f^\nu_B - f^\nu_A f^\mu_B), \\
R_{\lambda\mu} (\Omega ) & = & \dagOm_\lambda (T^{\rm T}_\lambda \dagOm_\mu) (T^{\rm T}_\mu \Omega_\lambda) \Omega_\mu .                                 
\end{eqnarray}

\noindent Here $T_\lambda$ is translation operator along the edge $\lambda$ to the neighboring site; $f^\lambda_A$ (or $f_{\lambda A}$) and $\Omega^{AB}_\lambda$ are freely chosen vector and matrix field variables at the sites (vertices), that is, in hypercubes. The $d$-vector will be denoted in bold, like $\vf_\lambda$ for $f_{\lambda A}$. Assuming this periodic structure, we are able to analyze different areas in a unified manner in the same notation. At the same time, due to possibility of discontinuous $g_{\lambda \mu} = f^A_\lambda f_{\mu A}$, using hypercubic decomposition instead of a general simplicial decomposition makes no restrictions on the form of the metric, which can be approximated (in a stepwise manner) by a set of hypercubes.

Here
\begin{equation}                                                            
\dagOm = (\Omega \eta)^{\rm T} \eta, ~~~ \eta = {\rm diag} (+1, \dots, +1, -1).
\end{equation}

\noindent The matrix $\Omega$ as an element of SO(d-1,1) by default has one upper and one lower index, and $\eta$ serves to transform these two to the same level and vice versa.

The second term in Eq. (\ref{S-discr}) represents an additional condition on $\Omega_\lambda$, multiplied by the Lagrange multiplier $\Lambda^\lambda_{[\mu \nu]}$. This condition violates the local gauge SO(d-1,1) symmetry. This circumstance does not seem to be a critical shortcoming of the formalism (the main motivation for the introduction of gauge theories was their renormalizability, but in the case of gravity the Cartan-Weyl action itself is not renormalizable in spite of the local gauge symmetry).

The term with Barbero-Immirzi parameter $\gamma$ in the Faddeev gravity \cite{I} can be easily incorporated into discrete formalism, to give, in addition to the Eq. (\ref{S-discr}) the term
\begin{eqnarray}                                                            
& & S^{\rm discr}_\gamma = \frac{1}{8 \pi G \gamma} \sum_{\rm sites} \sum_{\lambda, \mu} \frac{\sqrt{ (\epsilon^{\lambda \mu \nu \rho} f_{\nu A} f_{\rho B} )^2}}{2 } \arcsin \left[ \frac{\epsilon^{\lambda \mu \nu \rho} f_{\nu A} f_{\rho B} }{2 \sqrt{ (\epsilon^{\lambda \mu \nu \rho} f_{\nu A} f_{\rho B}  )^2}} R^{AB}_{\lambda\mu} (\Omega ) \right], \\
& & \epsilon^{0123} = +1. \nonumber
\end{eqnarray}

Let us pass to the continuous time. This procedure assumes that the 4-dimensional piecewise flat manifold is constructed of 3-dimensional piecewise flat manifolds (leaves) of a similar structure labeled by a parameter $t$ (time), so that $t$ and $t \pm \d t$ label neighboring leaves and $\d t \to 0$. It is also assumed that
\begin{equation}                                                            
f^A_0 = O( \d t), ~~~ \Omega_0 = 1 + O( \d t)
\end{equation}

\noindent and the functions considered are differentiable so that
\begin{equation}                                                            
T_0 = 1 + \d t \frac{\d }{\d t} + O((\d t)^2).
\end{equation}

We have
\begin{equation}                                                            
R_{0 \lambda} - 1 = - ( R_{\lambda 0} - 1 ) = \dagOm_\lambda \dotOm_\lambda \d t + \dots .
\end{equation}

\noindent Therefore the full action
\begin{eqnarray}                                                           
& & \hspace{-10mm} S^{\rm discr} + S^{\rm discr}_\gamma = \nonumber \\ & & \hspace{-10mm} \int \frac{\d t}{16 \pi G} \sum_{\rm sites} \sum_\lambda \left [ \sqrt{-\det \| g_{\lambda \mu} \|} (f^0_A f^\lambda_B - f^\lambda_A f^0_B) + \frac{1}{\gamma} \epsilon^{0 \lambda \mu \nu} f_{\mu A} f_{\nu B} \right ] (\dagOm_\lambda \dotOm_\lambda)^{AB} + \dots .
\end{eqnarray}

\noindent Consider here contribution of certain quadrangle, say, that one formed by $\vf_1$ and $\vf_2$ at certain vertex. Let $\vn_1$, $\vn_2$ be a pair of mutually orthogonal unit vectors in the plane of $\vf_1$, $\vf_2$. Then
\begin{equation}                                                           
[\vf_1 \vf_2]_{AB} \equiv f_{1A} f_{2B} - f_{2A} f_{1B} = A [\vn_1 \vn_2]_{AB}
\end{equation}

\noindent where $A$ is the area of the quadrangle. Use for a moment the following identity,
\begin{equation}                                                           
( f^\lambda_A f^\mu_B - f^\mu_A f^\lambda_B )\sqrt{-g} = - \frac{1}{2} \epsilon^{\lambda \mu \nu \rho} \epsilon_{A B C D} f^C_\nu f^D_\rho,
\end{equation}

\noindent where $\epsilon_{A B C D}$ is the completely antisymmetric tensor in the horizontal subspace,
\begin{equation}                                                           
\epsilon_{A B C D} = \frac{\epsilon^{\lambda \mu \nu \rho} f_{\lambda A} f_{\mu B} f_{\nu C} f_{\rho D}}{\sqrt{-g}}.
\end{equation}

\noindent Then
\begin{equation}                                                           
\epsilon_{A B C D} n^C_1 n^D_2 = [\vn_0 \vn_3]_{AB}
\end{equation}

\noindent where $\vn_0, \vn_1, \vn_2, \vn_3$ form an orthogonal frame in the horizontal subspace,
\begin{equation}                                                           
\vn_\lambda \cdot \vn_\mu = 0 \mbox{~~~at~~~} \lambda \neq \mu, ~~~ \vn^2_1 = \vn^2_2 = \vn^2_3 = +1, ~~~ \vn^2_0 = -1.
\end{equation}

\noindent With these notations, contribution of the considered quadrangle to the kinetic term in the Lagrangian takes the form
\begin{equation}\label{L-kin}                                              
L = \frac{A}{16 \pi G} {\rm tr} \{ ( - \frac{1}{\gamma} [\vn_1 \vn_2] + [\vn_0 \vn_3] ) \eta \dagOm_3 \dotOm_3 \}.
\end{equation}

\noindent Here $\dagOm_3 \dotOm_3$ has one upper and one lower index, and $\eta$ serves to transform these two to the same level, just as the antisymmetric bi-vector $[\vn_\lambda \vn_\mu]$ has.

Next we need an ansatz for $\Omega$. In the continuum theory we have
\begin{equation}\label{Faddeev-Cartan-Weyl}                                
S = \int [ R_{\lambda \mu}^{AB} (\omega ) + \Lambda^\nu_{[\lambda \mu]} \omega_{\nu}^{AB} ] f^\lambda_A f^\mu_B \sqrt{- g} \d^4 x
\end{equation}

\noindent for the first order representation of the Faddeev formulation of gravity \cite{I}. Here $\Lambda^\nu_{[\lambda \mu]}$ are the Lagrange multipliers. The latter are the coefficients at the constraints stating vanishing horizontal-horizontal sub-matrix of the connection matrix $\omega_{\nu}^{AB}$. Then it is only horizontal-vertical (vertical-horizontal) block of $\omega_{\nu}^{AB}$ which contributes to the action, because contribution of the vertical-vertical block identically vanishes. In the discrete formulation, the vertical-vertical part of $\Omega$ contributes nonzero value to the action due to the more complicated nonlinear structure of the latter. Therefore there is an ambiguity when passing to the discrete theory from the continuum formulation depending on whether we impose some constraints on the vertical-vertical components of the connection or not. We choose maximally simple ansatz issuing from the continuum theory with zero vertical-vertical part of $\omega_{\nu}^{AB}$. That is, in overall,
\begin{equation}                                                           
\omega_{\nu}^{AB} f^\lambda_A f^\mu_B = 0, ~~~ \omega_{\nu}^{AB} \Pi_{AC} \Pi_{BD} = 0.
\end{equation}

\noindent This has the general solution
\begin{equation}\label{omega=fb}                                           
\omega_{\lambda AB} = f^\mu_A b_{\mu \lambda B} - f^\mu_B b_{\mu \lambda A} \equiv [ \vf^\mu \vb_{\mu \lambda} ]_{AB}, ~~~ \vb_{\mu \lambda } \cdot \vf_\nu = 0
\end{equation}

\noindent parameterized by a new independent vertical vector variable $\vb_{\mu \lambda}$. Upon substituting this to the action we can find $b_{\mu \lambda A}$ from the equations of motion for it,
\begin{equation}                                                           
b_{\mu \lambda A} = f_{\mu B, \lambda} \Pi^B_A.
\end{equation}

\noindent We can view $\vf^\mu$ in (\ref{omega=fb}) being expanded over a set of the orthogonal frame vectors $\vn_\mu$ and get analogous expression for $\omega_\lambda$ with some redefined vertical vector variable $\tilde{\vb}_{\mu \lambda}$,
\begin{equation}                                                           
\omega_{\lambda} = \sum_\mu [ \vn_\mu \tilde{\vb}_{\mu \lambda} ] \eta .
\end{equation}

\noindent Now omitting tildes we can transfer this to the discrete case and $\Omega_3$ of interest,
\begin{equation}\label{Omega-dagOmega=nb}                                  
\frac{1}{2} ( \Omega_3 - \dagOm_3 ) = \sum_\lambda [ \vn_\lambda \vb_\lambda ] \eta + \dots .
\end{equation}

\noindent Here the dots reflect certain freedom in the choice of the exact form of the constraints in Eq. (\ref{S-discr}). Namely, according to the status of the discrete form of the Faddeev action \cite{II}, this form should reproduce the continuum action when there is certain fixed continuum (smooth) distribution of $f^\lambda_A$ on the fixed smooth manifold, and the considered piecewise flat geometry is only approximation to this continuum one which is made more and more fine by tending typical triangulation length to zero. So the dots mean the terms which do not contribute to the continuum limiting action. Let $\delta f$ be typical variation of $f^\lambda_A$ when passing from simplex to simplex. The Eq. (\ref{Omega-dagOmega=nb}) has the order of magnitude $O(\delta f)$, and the dots a'priori should have the order of magnitude $O((\delta f)^2)$ (more accurately, $O((\delta f)^3)$ with taking into account SO group properties of $\Omega_3$). Using this freedom, we can restore full nonlinear structure of $\Omega_3$ as
\begin{equation}\label{Omega=uuuu}                                         
\Omega_3 = u_2 u_0 u_3 u_1, ~~~ u_\lambda = \exp ( [ \vn_\lambda \vb_\lambda ] \eta ).
\end{equation}

\noindent Despite of the non-commutativity of the different matrices $u$, the RHS of Eq. (\ref{Omega=uuuu}) can be rewritten with another order of sequence of the subscripts 0, 1, 2, 3, although with appropriate redefinition of $\vb$'s,
\begin{eqnarray}\label{Omega=uuuu-new}                                     
& & \Omega_3 = u_{02} u_2 u_1 u_{31^+}, \\ & & u_{02} = \exp ( [ \vn_0 \vb_{02} ] \eta ), ~~~ \vb_{02} = u_2 \vb_0, \nonumber\\ & & u_{31^+} = \exp ( [ \vn_3 \vb_{31^+} ] \eta ), ~~~ \vb_{31^+} = \dagu_1 \vb_3. \nonumber
\end{eqnarray}

Now substitute Eq (\ref{Omega=uuuu}) to the $1/\gamma$ term in (\ref{L-kin}), and Eq (\ref{Omega=uuuu-new}) to the primary (purely Cartan-Weyl) second term. Besides that, let us parameterize the unit area bi-vector not as $[ \vn_1 \vn_2 ]$, but as $\dagu_1 [ \vn_1 \vn_2 ] u_1$, and let us parameterize the dual unit area bi-vector not as $[ \vn_0 \vn_3 ]$, but as $\dagu_{31^+} [ \vn_0 \vn_3 ] u_{31^+}$. This is possible because these newly parameterized bi-vectors are nothing but rotated by $\dagu_1 u_3$ the bi-vectors $[ \vn_1 \vn_2 ]$ and $[ \vn_0 \vn_3 ]$ and therefore these are mutually dual ones. Simple calculation gives
\begin{equation}\label{L-kin-ansatz}                                       
\frac{16 \pi G L}{A} = - \frac{1}{\gamma} {\rm tr} [ \vn_1 \vn_2] \eta (\dagu_2 \dotu_2 + \dotu_1 \dagu_1 ) + {\rm tr} [\vn_0 \vn_3] \eta ( \dagu_{02} \dotu_{02} + \dotu_{31^+} \dagu_{31^+} ).
\end{equation}

\noindent The exact nonlinear form of the adopted ansatz for connection is quite simple, e. g. at $\vn^2 = +1$
\begin{eqnarray}                                                           
& & \exp ([ \vn \vb ] \eta ) \eta = \eta + ( \vn \otimes \vn + \vl \otimes \vl ) ( \cos b - 1) + [\vn \vl ] \sin b, \\ & & b = \sqrt{\vb^2}, ~~~ \vl = \vb / b. \nonumber
\end{eqnarray}

\noindent This allows to rewrite Eq. (\ref{L-kin-ansatz}) as
\begin{eqnarray}\label{L-kin-final}                                        
\frac{8 \pi G L}{A} = - \frac{1}{\gamma} \left [ (\vn_2 \cdot \dotvn_1) (\cos b_2 - \cos b_1 ) + (\vl_1 \cdot \dotvn_2 ) \sin b_1 - (\vl_2 \cdot \dotvn_1 ) \sin b_2 \right ] \nonumber \\ - ( \vn_0 \cdot \dotvn_3 ) ( \cosh b_0 - \cos b_3 ) - ( \vl_{31^+} \cdot \dotvn_0 )\sin b_3 - (\vl_{02} \cdot \dotvn_3 ) \sinh b_0.
\end{eqnarray}

\section{ Elementary area spectrum}

One consequence of the additional conditions on $\Omega_\lambda$ (see Eq. (\ref{S-discr})) and, therefore, on $\vb_\lambda$, $\vb_\lambda \cdot \vn_\mu = 0$ (verticality), lies in the fact that the time derivative of $b_\lambda$ is not determined from the equations of motion (see Eq. (\ref{L-kin-final})). That is, $b_\lambda$ are non-dynamical. Now, performing a series of elementary transformations of the variables $\vb$, $\vn$, we find the range of the non-dynamic variables, where the spectra of area $A$, defined from requirements that wave function of different angle type variables $q$ be single-valued, are compatible with each other. In addition, this spectrum will be universal for all areas, up to the addition of terms vanishing in the continuum limit, to the discrete Lagrangian.

First consider the purely $1/\gamma$-part. Let us represent $\vl_1$, $\vl_2$ in terms of orthogonal vectors,
\begin{eqnarray}                                                           
& & \vl_1 = \lambda_1 \ve_1 + \lambda_2 \ve_2, ~~~ \vl_2 = \lambda_1 \ve_1 - \lambda_2 \ve_2, \\ & & \ve_1^2 = \ve_2^2 = 1, ~~~ \ve_1 \cdot \ve_2 = 0, ~~~ \lambda_{1,2} = \sqrt{\frac{1 \pm \vl_1 \cdot \vl_2 }{2}}. \nonumber
\end{eqnarray}

\noindent Thus we arrive at a combination of terms $\vn_2 \cdot \dotvn_1$, $\ve_1 \cdot \dotvn_2$, $\ve_2 \cdot \dotvn_2$, $\ve_1 \cdot \dotvn_1$, $\ve_2 \cdot \dotvn_1$. Let us pass to the new field variables by rotating in the 2-planes of $\vn_1, \vn_2$ and $\ve_1, \ve_2$,
\begin{equation}                                                           
\left. \begin{array}{c}
\vn_1 = \prvn_1 \cos \alpha - \prvn_2 \sin \alpha \\
\vn_2 = \prvn_1 \sin \alpha + \prvn_2 \cos \alpha
\end{array} \right\} , ~~~
\left. \begin{array}{c}
\ve_1 = \phantom{-} \prve_1 \cos \beta + \prve_2 \sin \beta \\
\ve_2 = - \prve_1 \sin \beta + \prve_2 \cos \beta
\end{array} \right\} .
\end{equation}

\noindent Let us choose $\alpha, \beta$ to cancel the terms with $\prve_1 \cdot \prdotvn_1$ and $\prve_2 \cdot \prdotvn_2$. Next rotate in the 2-planes of $\prvn_1, \prve_1$ and $\prvn_2, \prve_2$,
\begin{equation}                                                           
\left. \begin{array}{c}
\prvn_i = \pprvn_i \cos \alpha_i - \pprve_i \sin \alpha_i \\
\prve_i = \pprvn_i \sin \alpha_i + \pprve_i \cos \alpha_i
\end{array} \right\} i = 1, 2,
\end{equation}

\noindent choosing $\alpha_1, \alpha_2$ to cancel the terms with $\pprve_2 \cdot \pprdotvn_1$ and $\pprve_1 \cdot \pprdotvn_2$. The resulting $1/\gamma$-part $L_\gamma$ of the kinetic term $L$ reads
\begin{eqnarray}\label{L-gamma}                                            
& & - \gamma \frac{8 \pi G}{A} L_\gamma = \frac{(\cos b_2 - \cos b_1)(\sin^2 b_1 - \sin^2 b_2 )^2}{(\sin^2 b_1 - \sin^2 b_2 )^2 + 4 (\vl_1 \cdot \vl_2 )^2 \sin^2 b_1 \sin^2 b_2 } ~ \frac{\d}{\d t} ~ \frac{ (\vl_1 \cdot \vl_2 ) \sin b_1 \sin b_2 }{ \sin^2 b_1 - \sin^2 b_2 } \nonumber \\
& & \phantom{- \gamma \frac{8 \pi G}{A} L =} + \frac{1}{2} \left ( \sqrt{2 - 2 \cos b_1 \cos b_2 + 2 \sin \llangle \sin b_1 \sin b_2} \right. \nonumber \\ & & \phantom{- \gamma \frac{8 \pi G}{A} L =} \left. \phantom{\frac{1}{2} \left ( \left ( \right ) \right )} + \sqrt{2 - 2 \cos b_1 \cos b_2 - 2 \sin \llangle \sin b_1 \sin b_2} \right ) \pprvn_2 \cdot \pprdotvn_1 \nonumber \\
& & \phantom{- \gamma \frac{8 \pi G}{A} L =} + \frac{1}{2} \left ( \sqrt{2 - 2 \cos b_1 \cos b_2 + 2 \sin \llangle \sin b_1 \sin b_2} \right. \nonumber \\ & & \phantom{- \gamma \frac{8 \pi G}{A} L =} \left. \phantom{\frac{1}{2} \left ( \left ( \right ) \right )} - \sqrt{2 - 2 \cos b_1 \cos b_2 - 2 \sin \llangle \sin b_1 \sin b_2} \right ) \pprve_2 \cdot \pprdotve_1
\end{eqnarray}

\noindent Here $\llangle$ is the angle between $\vl_1$ and $\vl_2$. If we would like to describe the quantum state by the functions of the (angle type) coordinates of either $\pprvn_2, \pprvn_1$ or $\pprve_2, \pprve_1$, the $A$ should have a discrete spectrum. If we would like to describe the quantum state by the functions of the coordinates of all four $\pprvn_2, \pprvn_1, \pprve_2, \pprve_1$, the requirement of the consistency of spectrum defined from $\pprvn_2 \cdot \pprdotvn_1$ and from $\pprve_2 \cdot \pprdotve_1$ terms leads to
\begin{equation}                                                           
\sin \llangle \sin b_1 \sin b_2 = 0.
\end{equation}

\noindent The solution $\sin \llangle = 0$ fixes $\vl_2 = \pm \vl_1$, that is, it freezes $d - 1$ degrees of freedom. The solution $b_1 = 0$ (or $b_2 = 0$) means $\vb_1 = 0$ (or $\vb_2 = 0$) and thus fixes $d$ degrees of freedom. Finally, the solution $b_1 = \pi$ (or $b_2 = \pi$) cancels only one degree of freedom. Thus, if compared to the latter solution, the former two describe hypersurface in the configuration superspace of zero measure and should be discarded from the probabilistic consideration. So let us take $b_1 = \pi$ for definiteness.

Next consider the purely $(\gamma)^0$ part. Let us represent $\vl_\3$, $\vl_\0$ in terms of orthogonal vectors,
\begin{eqnarray}                                                           
& & \vl_\3 = \lambda_3 \ve_3 + \lambda_0 \ve_0, ~~~ \vl_\0 = \lambda_3 \ve_3 - \lambda_0 \ve_0, \\ & & \ve_3^2 = \ve_0^2 = 1, ~~~ \ve_3 \cdot \ve_0 = 0, ~~~ \lambda_{3,0} = \sqrt{\frac{1 \pm \vl_\3 \cdot \vl_\0 }{2}}. \nonumber
\end{eqnarray}

\noindent Thus we arrive at a combination of terms $\vn_0 \cdot \dotvn_3$, $\ve_3 \cdot \dotvn_0$, $\ve_0 \cdot \dotvn_0$, $\ve_3 \cdot \dotvn_3$, $\ve_0 \cdot \dotvn_3$. Let us rotate in the 2-planes of $\vn_3, \vn_0$ and $\ve_3, \ve_0$,
\begin{equation}                                                           
\left. \begin{array}{c}
\vn_3 = \prvn_3 \cosh \xi + \prvn_0 \sinh \xi \\
\vn_0 = \prvn_1 \sinh \xi + \prvn_0 \cosh \xi
\end{array} \right\} , ~~~
\left. \begin{array}{c}
\ve_3 = \phantom{-} \prve_3 \cos \zeta + \prve_0 \sin \zeta \\
\ve_0 = - \prve_3 \sin \zeta + \prve_0 \cos \zeta
\end{array} \right\} .
\end{equation}

\noindent Let us choose $\zeta, \xi$ to cancel the terms with $\prve_3 \cdot \prdotvn_3$ and $\prve_0 \cdot \prdotvn_0$. Next rotate in the 2-planes of $\prvn_3, \prve_3$ and $\prvn_0, \prve_0$,
\begin{equation}                                                           
\left. \begin{array}{c}
\prvn_3 = \pprvn_3 \cos \alpha_3 - \pprve_3 \sin \alpha_3 \\
\prve_3 = \pprvn_3 \sin \alpha_3 + \pprve_3 \cos \alpha_3
\end{array} \right\} , ~~~
\left. \begin{array}{c}
\prvn_0 = \pprvn_0 \cosh \xi_0 + \pprve_0 \sinh \xi_0 \\
\prve_0 = \pprvn_0 \sinh \xi_0 + \pprve_0 \cosh \xi_0
\end{array} \right\} ,
\end{equation}

\noindent choosing $\alpha_3, \xi_0$ to cancel the terms with $\pprve_0 \cdot \pprdotvn_3$ and $\pprve_3 \cdot \pprdotvn_0$. The resulting $(\gamma )^0$-part $L_0$ of the kinetic term $L$ reads
\begin{eqnarray}\label{L-0}                                                
& & - \frac{8 \pi G}{A} L_0 = \frac{(\cosh b_0 - \cos b_3)(\sinh^2 b_0 + \sin^2 b_3 )^2}{(\sinh^2 b_0 + \sin^2 b_3 )^2 - 4 (\vl_\3 \cdot \vl_\0 )^2 \sin^2 b_3 \sinh^2 b_0 } ~ \frac{\d}{\d t} ~ \frac{ (\vl_\3 \cdot \vl_\0 ) \sin b_3 \sinh b_0 }{ \sinh^2 b_0 + \sin^2 b_3 } \nonumber \\
& & \phantom{- \frac{8 \pi G}{A} L =} + \frac{1}{2} \left ( \sqrt{2 - 2 \cos b_3 \cosh b_0 + 2 i \sin \nllangle \sin b_3 \sinh b_0} \right. \nonumber \\ & & \phantom{- \frac{8 \pi G}{A} L =} \left. \phantom{\frac{1}{2} \left ( \left ( \right ) \right )} + \sqrt{2 - 2 \cos b_3 \cosh b_0 - 2 i \sin \nllangle \sin b_3 \sinh b_0} \right ) \pprvn_0 \cdot \pprdotvn_3 \nonumber \\
& & \phantom{- \frac{8 \pi G}{A} L =} + \frac{1}{2 i} \left ( \sqrt{2 - 2 \cos b_3 \cosh b_0 + 2 i \sin \nllangle \sin b_3 \sinh b_0} \right. \nonumber \\ & & \phantom{- \frac{8 \pi G}{A} L =} \left. \phantom{\frac{1}{2} \left ( \left ( \right ) \right )} - \sqrt{2 - 2 \cos b_3 \cosh b_0 - 2 i \sin \nllangle \sin b_3 \sinh b_0} \right ) \pprve_0 \cdot \pprdotve_3 .
\end{eqnarray}

\noindent If we would like to describe the quantum state by the functions of the (angle type) coordinates of $\pprve_0, \pprve_3$, the $A$ should have a discrete spectrum. (The $\pprvn_0 \cdot \pprdotvn_3$ term does not lead to the discrete spectrum since $\pprvn_0$ varies in the noncompact region.) Consistency with the above description also by the functions of the coordinates of $\pprvn_2, \pprvn_1, \pprve_2, \pprve_1$ requires
\begin{equation}                                                           
\sin \nllangle \sin b_3 \sinh b_0 = 0.
\end{equation}

\noindent Again, the solution $b_3 = \pi$ provides us with the configuration subspace of the largest dimensionality.

Typically, if the considered piecewise flat geometry is seen on sufficiently large scale as some smooth geometry with a typical curvature $R$, the angle defect has the order of magnitude $a^2R$ where $a$ is a typical elementary (triangulation) length. We have found by evaluating area/length path integral  probability distribution in the theory of the similar type (based on exact connection representation of the Regge action) that we can take $a = l_p$ \cite{kha}, the Planck scale. If $\rho$ is the typical energy density, then $R = l^2_p \rho$. The resulting angle defect $l^4_p \rho$ turns out to be extremely small for all ordinary types of matter. Evidently, this can correspond to the small $\vb_\lambda$ as well, that is, to $\vb_\lambda$ in some small neighborhood of zero.

Above we have found that the consistent quantum description in terms of the considered variables is achieved at the non-perturbative points when some $\vb_\lambda$ are $\pi$ in absolute value (describe reflections as factors in $\Omega$). Fortunately, the curvature describing small neighborhood of the flat space-time can be reproduced using reflections in the connection matrices as well. That is, we can consider the above found points like $(b_1, b_2, b_3, b_0) = (\pi, \varepsilon_2, \pi, \varepsilon_0)$ with small $\varepsilon_\lambda$.

Of course, the question arises whether we can relax the above established restrictions of the type of $b_1 = \pi, b_3 = \pi$. If $(b_1, b_2, b_3, b_0) = (\pi + \varepsilon_1, \varepsilon_2, \pi + \varepsilon_3, \varepsilon_0)$, then the coefficients at $\pprve_2 \cdot \pprdotve_1$ and $\pprve_0 \cdot \pprdotve_3$ in Eqs. (\ref{L-gamma}) and (\ref{L-0}) have the order of $\varepsilon^2$. Again, consider a fixed continuum geometry with typical curvature $R$ in the given region, and let the considered piecewise flat geometry is just triangulation of this continuum one which is made more and more fine by tending typical elementary (triangulation) length $a$ to zero. Then the order of magnitude of the actual values of $\varepsilon^2$ is bounded from above by $a^2 R$. The contribution of $\pprve_2 \cdot \pprdotve_1$ and $\pprve_0 \cdot \pprdotve_3$ to $L$ are of the order of $A a^2 R \sim a^4 R$. Since the number of cubes with edge length $a$ in the given volume $V$ in the 3-dimensional section is $V/a^3$, the contribution of these terms to $L$ in this volume has the order $O(a)$ and thus tends to zero in the continuum limit $a \to 0$. Besides these terms, the non-constant parts of the coefficients of $\pprvn_2 \cdot \pprdotvn_1$, $\pprvn_0 \cdot \pprdotvn_3$ are $O(\varepsilon^2)$. Thus, we can write
\begin{equation}                                                           
L = - \frac{A}{4 \pi G} \left ( \frac{1}{\gamma} \pprvn_2 \cdot \pprdotvn_1 + \pprvn_0 \cdot \pprdotvn_3 \right ) + O(\varepsilon^2).
\end{equation}

\noindent Here $O(\varepsilon^2)$ terms can be canceled by modifying discrete action by adding certain terms to it vanishing in the continuum limit. Further, whereas $\pprvn_2 \cdot \pprvn_1 = 0 = \pprvn_0 \cdot \pprvn_3$, the scalar products of $\pprvn_2, \pprvn_1$ with $\pprvn_0, \pprvn_3$ are $O(\varepsilon^2)$. We can orthogonalize the set $\pprvn_\lambda$ and get orthogonal normalized $\tvn_\lambda$ so that
\begin{equation}                                                           
\pprvn_\lambda = \tvn_\lambda + O(\varepsilon^2),
\end{equation}

\noindent and
\begin{equation}                                                           
L = - \frac{A}{4 \pi G} \left ( \frac{1}{\gamma} \tvn_2 \cdot \dottvn_1 + \tvn_0 \cdot \dottvn_3 \right ) + O(\varepsilon^2)
\end{equation}

\noindent where $\tvn_\lambda$ up to $O(\varepsilon^2)$ are linear combinations of $\vn_\lambda$ and $\vvarep_\lambda = \varepsilon_\lambda \vl_\lambda$, for example,
\begin{equation}                                                           
\tvn_1 = \vn_1 + \frac{1}{2} \varepsilon_1 \vl_1 + O(\varepsilon^2) = \vn_1 + \frac{1}{2} \left ( 1 - \frac{\pi}{b_1} \right ) \vb_1 + O(\varepsilon^2).
\end{equation}

Thus, if we allow the modification of the discrete action (adding terms that vanish in the continuum limit), the kinetic term can be the following (the leading part of $L$, which does not depend on non-dynamic variables),
\begin{equation}                                                           
\tilde{L} = - \frac{A}{4 \pi G} \left ( \frac{1}{\gamma} \tvn_2 \cdot \dottvn_1 + \tvn_0 \cdot \dottvn_3 \right )
\end{equation}

\noindent (as far as the area in the coordinates $x^1, x^2$ is considered). Defining area spectrum is then straightforward. Consistent dependence of the wave function on the angle coordinates of the vectors $\tvn_0, \tvn_3$ does not impose requirement of the discreteness of area spectrum since their relative coordinates vary in the noncompact region. However, these vectors occupy 2-dimensional subspace and thus leave $d-2$ dimensions for $\tvn_2, \tvn_1$. Let us use for $\tvn_1$ in the $(d-2)$-dimensional subspace the spherical coordinates $\chi_1, \dots, \chi_{d-3}$,
\begin{equation}                                                           
\tvn_1 = \left ( \begin{array}{l}
\sin \chi_{d-3} \sin \chi_{d-4} \dots \sin \chi_1 \\
\sin \chi_{d-3} \sin \chi_{d-4} \dots \cos \chi_1 \\
\vdots \\
\cos \chi_{d-3}
\end{array} \right ).
\end{equation}

\noindent The basis in the $(d-3)$-dimensional orthogonal subspace is naturally chosen as
\begin{eqnarray}                                                           
& & \tvn^{(1)}_2 = \left ( \begin{array}{l}
\phantom{-} \cos \chi_1 \\
- \sin \chi_1 \\
\phantom{-} 0 \\
\phantom{-} \vdots \\
\phantom{-} 0
\end{array} \right ) , ~~~
\tvn^{(2)}_2 = \left ( \begin{array}{l}
\phantom{-} \cos \chi_2 \sin \chi_1 \\
\phantom{-} \cos \chi_2 \cos \chi_1 \\
- \sin \chi_2 \\
\phantom{-} \vdots \\
\phantom{-} 0
\end{array} \right ) , \dots , \nonumber \\
& & \tvn^{(d-3)}_2 = \left ( \begin{array}{l}
\phantom{-} \cos \chi_{d-3} \sin \chi_{d-4} \dots \sin \chi_1 \\
\phantom{-} \cos \chi_{d-3} \sin \chi_{d-4} \dots \cos \chi_1 \\
\phantom{-} \vdots \\
\phantom{-} \vdots \\
- \sin \chi_{d-3}
\end{array} \right ).
\end{eqnarray}

\noindent Then we can write
\begin{equation}                                                           
\tvn_2 = C_1 \tvn^{(1)}_2 + \dots +C_{d-3} \tvn^{(d-3)}_2, ~~~ C^2_1 + \dots +C^2_{d-3} = 1.
\end{equation}

\noindent Substituting these expressions to $\tilde{L}$ we find the $\dot{\chi}$-terms in $\tilde{L}$,
\begin{equation}                                                           
\tilde{L}_{\dot{\chi}} = - \frac{A}{4 \pi G \gamma} \left (C_1 \dot{\chi}_1 \sin \chi_{d-3} \dots \sin \chi_2 + \dots + C_{d-3} \dot{\chi}_{d-3} \right ).
\end{equation}

\noindent The $C_1, \dots, C_{d-3}$ and $A$ can serve to parameterize the conjugate to $\chi_1, \dots, \chi_{d-3}$ momenta $p_n = \partial \tilde{L} / \partial \dot{\chi}_n$. In particular,
\begin{equation}\label{AA}                                                 
\left ( \frac{A}{4 \pi G \gamma} \right )^2 = \left (\frac{p_1}{\sin \chi_{d - 3} \dots \sin \chi_2} \right )^2 + \cdots +p^2_{d - 3}.
\end{equation}

\noindent In quantum theory $p_n$ are substituted by the operators $-i\partial /\partial \chi_n$, and under appropriate product ordering the RHS of Eq. (\ref{AA}) is nothing but minus angle part of the $(d - 2)$-dimensional Laplace operator. The spectrum of the latter is well-known (see, e.g., Ref. \cite{Vilen}). Its eigenfunctions are labeled by $d-3$ integers $(j, k_1, \dots, \pm k_{d-4})$ such that $j \geq k_1 \geq \dots \geq k_{d-4} \geq 0$. The eigenvalues $j(j + d - 4)$ depend on $j$ only. Then the number of eigenfunctions is
\begin{equation}                                                           
g(j) = \frac{(j + d - 5)!(2j + d - 4)}{j!(d - 4)!}
\end{equation}

\noindent for the $j$-th eigenvalue, that is, for the area
\begin{equation}                                                           
A \equiv 8 \pi l^2_p \gamma a(j), ~~~ a(j) = \frac{1}{2} \sqrt{j (j + d - 4)}, ~~~ l^2_p = G.
\end{equation}

In the calculation of Ref \cite{Khr2}, $g(j)$ is the statistical weight of elementary areas $8 \pi l^2_p \gamma a(j)$ with quantum number $j$. In this calculation, the requirement that the formula for statistical entropy would coincide with the Bekenstein-Hawking relation (which states that the entropy of the black hole is $(4 l^2_p)^{-1} A_{bh}$ where $A_{bh}$ is the horizon area) gives
\begin{equation}                                                           
\sum_j g(j) e^{-2 \pi \gamma a(j)} = 1.
\end{equation}

\noindent This relation is an equation for $\gamma$. Solving it for the genuine Faddeev's choice of dimensionality of the external space $d = 10$ we find
\begin{equation}                                                           
\gamma = 0.393487933....
\end{equation}

\noindent In principle, one can consider taking another $d$. Note that at $d < 10$ there is difficulty in reducing the classical Faddeev action to the Einstein one because the number of the vertical components of the equations of motion $4 (d - 4)$ is not enough to provide vanishing 24 components of the torsion $T^\lambda_{[\mu \nu ]}$ necessary for that. As for $d > 10$, such dimensionality might be of interest when considering {\it global} embedding into the external Euclidean/Minkowskian space, namely, such embedding for the 4-dimensional spacetime may require as much as $d = 230$ dimensions \cite{230}. For this $d$ calculation gives
\begin{equation}                                                           
\gamma = 0.359772297....
\end{equation}

\noindent The dependence on $d$ is rather weak.

\section{ Conclusion}

The area spectrum arising in the considered formulation is physically reasonable, quantum being of the order of Planck scale $l^2_p$. Though, a priori this circumstance is not quite evident since the spectrum depends on the non-dynamic (that is, whose time derivative is indeterminate from the equations of motion) variables entering as parameters. Namely, this fact certainly does not hold in the usual perturbative framework around flat background when connection variables $\Omega$ are in the neighborhood of unity and thus $\vb$ in Eq. (\ref{L-kin-final}) are close to zero. Then the area spectrum is scaled by an infinite factor. However, studying the different requirements to the area spectrum (due to requirements that wave function be single-valued w. r. t. the different angle type variables $q$), we have found that the compatibility of these requirements is achieved mainly at the non-perturbative values of certain variables of the type of $b = \pi$.

Eventually, this can be traced back to the non-perturbative nature of the Faddeev gravity itself. Indeed, vertical equations of the continuum Faddeev gravity establishing its equivalency to the Einstein gravity on classical level are degenerate at $b_{\lambda \mu A} = 0$. This equality means flat space-time. The discrete counterparts $\vb_\lambda$ can provide flat space-time also when these have the lengths $\pi$.

Even if we have physically reasonable area spectrum, the dependence on the non-dynamical variables would mean that the spectrum is not universal for all areas. However, as it turned out, this spectrum can be made consistent and universal in the {\it neighborhood} of $b = \pi$ for some $b$, if we allow the addition of terms to the discrete action, which tend to zero in the formal continuum limit.

A discrete quantization of separate elementary piece of area (triangle) can be observed also in the simplicial minisuperspace formulation of the usual Einstein gravity, and in the absence of the Barbero-Immirzi term ($1 / \gamma = 0$) only timelike area is quantized \cite{kha}. In this respect, the situation differs from that in the continuum theory based on Ashtekar variables and loop states where the spacelike area is quantized in the absence of the Barbero-Immirzi term in the action \cite{Ash}. At $1 / \gamma \neq 0$ we can get discrete spacelike area spectrum proportional to $\gamma$ \cite{kha}. The problem of analyzing the spectrum of the total surface area lies in the fact that its constituent triangles are not independent, so that this spectrum can not be found simply as the sum of the spectra of individual independent triangles.

The author is grateful to I. B. Khriplovich for stimulating long-standing interest in the problem of area spectrum and its black hole applications and valuable discussions. The author thanks I. A. Taimanov who had attracted author's attention to the Faddeev formulation of gravity and Ya.V. Bazaikin for discussions of this subject. The present work was supported by the Ministry of Education and Science of the Russian Federation, Russian Foundation for Basic Research through Grant No. 11-02-00792-a and Grant 14.740.11.0082 of federal program "personnel of innovational Russia".


\begin{thebibliography}{99}
\bibitem{Fad}
 L. D. Faddeev, New dynamical variables in Einstein's theory of gravity, {\it Theor. Math.
 Phys.} {\bf 166}, 279-290 (2011).
\bibitem{II}
 V. M. Khatsymovsky, Faddeev formulation of gravity in discrete form,
 arXiv: $\!\!\!\!$ 1201.0808[gr-qc].
\bibitem{I}
 V. M. Khatsymovsky, First order representation of the Faddeev formulation of gravity,
 arXiv: $\!\!\!\!$ 1201.0806[gr-qc].
\bibitem{Ash}
 A. Ashtekar, C. Rovelli and L. Smolin, Weaving a classical geometry with quantum threads, {\it Phys. Rev. Lett.} {\bf 69}, 237-240 (1992); arXiv:hep-th/9203079.
\bibitem{Loll}
 R. Loll, Further results on geometric operators in quantum gravity, {\it Class. Quant. Grav.} {\bf 14}, 1725-1741 (1997); arXiv:gr-qc/9612068.
\bibitem{ABCK}
 A. Ashtekar, J. Baez, A. Corichi and K. Krasnov, Quantum Geometry and Black Hole Entropy, {\it Phys.Rev.Lett.} {\bf 80}, 904-907 (1998); arXiv:gr-qc/9710007.
\bibitem{LewDom}
 J. Lewandowski and M. Domagala, Black hole entropy from Quantum Geometry, {\it Class. Quantum Grav.} {\bf 21}, 5233-5244 (2004); arXiv:gr-qc/0407051.
\bibitem{Mei}
 K. Meissner, Black hole entropy in Loop Quantum Gravity, {\it Class. Quantum Grav.} {\bf 21}, 5245-5252 (2004); arXiv:gr-qc/0407052.
\bibitem{Barb}
 J. F. Barbero, Real Ashtekar Variables for Lorentzian Signature Space-times, {\it Phys.
 Rev. D} {\bf 51}, 5507-5510 (1995); arXiv:gr-qc/9410014.
\bibitem{Imm}
 G. Immirzi, Quantum Gravity and Regge Calculus, {\it Nucl. Phys. Proc. Suppl.} {\bf 57}, 65-72 (1997); arXiv:gr-qc/9701052.
\bibitem{Holst}
 S. Holst, Barbero's Hamiltonian Derived from a Generalized Hilbert-Palatini Action,
 {\it Phys. Rev. D} {\bf 53}, 5966-5969 (1996); arXiv:gr-qc/9511026.
\bibitem{Fat}
 L. Fatibene, M. Francaviglia and C. Rovelli, Spacetime Lagrangian For\-mu\-la\-ti\-on of Barbero-Immirzi Gravity, {\it Class. Quantum Grav.} {\bf 24}, 4207-4218 (2007); arXiv:0706.1899.
\bibitem{Bek2}
 J. D. Bekenstein, Universal upper bound to entropy-to-energy ratio for bounded systems, {\it Phys. Rev. D} {\bf 23}, 287-298 (1981).
\bibitem{Tho}
 G. 't Hooft, Dimensional Reduction in Quantum Gravity, in {\it Salam Festschrift} (Singapore, 1993); arXiv:gr-qc/9310026.
\bibitem{Sus}
 L. Susskind, The World as a Hologram, {\it J. Math. Phys.} {\bf 36}, 6377-6396 (1995); arXiv:hep-th/9409089.
\bibitem{Khr1}
 I. B. Khriplovich and R. V. Korkin, How Is the Maximum Entropy of a Quantized Surface Related to Its Area?,
 {\it J. Exp. Theor. Phys.} {\bf 95}, 1-4 (2002); {\it Zh. Eksp. Teor. Fiz.} {\bf 95}, 5-9 (2002); arXiv:gr-qc/0112074.
\bibitem{Glo}
 A. Ghosh and P. Mitra, An improved estimate of black hole entropy in the quantum geometry approach, {\it Phys. Lett. B} {\bf 616}, 114-117 (2005); arXiv:gr-qc/0411035.
\bibitem{Khr2}
 I.B. Khriplovich, Quantized Black Holes, Their Spectrum and Radiation,
 {\it Phys. Atom. Nucl.} {\bf 71}, 671-680 (2008); arXiv:gr-qc/0506082.
\bibitem{Cor}
 A. Corichi, J. Diaz-Polo and E. Fernandez-Borja, Quantum geometry and microscopic black hole entropy, {\it Class. Quant. Grav.} {\bf 24}, 243-251 (2007); arXiv:gr-qc/0605014.
\bibitem{kha}
 V. M. Khatsymovsky, Integration over connections in the discretized gravitational functional integrals,
 {\it Mod. Phys. Lett. A} {\bf 25}, 351-368 (2010); arXiv:0912.1109.
\bibitem{Vilen}
 N. Ya. Vilenkin, {\it Special Functions and the Theory of Group
 Representations}, Translations of Mathematical Monographs, Vol. 22, (Amer.
 Math.  Soc., Providence, Rhode Island, 1968).
\bibitem{230}
 J.F. Nash, The imbedding problem for Riemannian manifolds,
 {\it Ann. Math.} {\bf 63}, 20-63 (1956).
\end{thebibliography}
\end{document}